\documentclass[runningheads]{llncs}

\usepackage{graphicx, subfigure, amsmath, amssymb, upgreek, array, graphbox,  verbatim, booktabs, siunitx, xcolor}

\newcommand{\mypar}[1]{\paragraph{\normalfont\textbf{#1}}}
\newcolumntype{C}[1]{>{\centering\let\newline\\\arraybackslash\hspace{0pt}}m{#1}}

\usepackage[colorinlistoftodos,prependcaption,textsize=tiny]{todonotes}

\newcommand{\matr}[1]{\mathbf{#1}}
\newcommand{\norm}[1]{\left\|#1\right\|}
\newcommand{\tr}[1]{{#1}^\top}

\newcommand{\yy}{\matr{y}}
\newcommand{\xx}{\matr{x}}
\newcommand{\Loss}{\mathcal{L}}
\newcommand{\Lwgan}{\Loss_{\mathrm{GAN}}}
\newcommand{\LwganX}{\Loss_{\mathrm{GAN}_X}}
\newcommand{\LwganY}{\Loss_{\mathrm{GAN}_Y}}
\newcommand{\Lcyc}{\Loss_{\mathrm{cyc}}}
\newcommand{\Id}{\mathrm{Id}}
\newcommand{\EE}{\mathbb{E}}

\newcommand{\logId}{\log_{\Id}}
\newcommand{\expId}{\exp_{\Id}}

\newcommand{\XHR}{X_{\mathrm{HR}}}
\newcommand{\YHR}{Y_{\mathrm{HR}}}
\newcommand{\YLR}{Y_{\mathrm{LR}}}
\newcommand{\genX}{G_X}
\newcommand{\genY}{G_Y}
\newcommand{\disX}{D_X}
\newcommand{\disY}{D_Y}
\newcommand{\mr}[1]{\mathrm{#1}}

\newcommand{\Prob}{\mathbb{P}}
\newcommand{\PP}{\matr{P}}

\newcommand{\UU}{\matr{U}}
\newcommand{\Eig}{\boldsymbol{\Upsigma}}

\newcommand{\Tid}{\mathcal{T}_{\Id}}
\newcommand{\Ss}{\matr{S}}

\newcommand{\ToneReal}[1]{\includegraphics[align=c,height=.240\textwidth,width=.245\textwidth]{figures/patches_t1/T1_real_#1.png}}
\newcommand{\ToneRec}[1]{\includegraphics[align=c,height=.240\textwidth,width=.245\textwidth]{figures/patches_t1/T1_rec_#1.png}}

\newcommand{\ToneRealFA}[1]{\includegraphics[height=.195\textwidth,width=.195\textwidth]{figures/patches_dti/fa_real_#1.PNG}}
\newcommand{\ToneGenFA}[1]{\includegraphics[height=.195\textwidth,width=.195\textwidth]{figures/patches_dti/fa_gen_#1.PNG}}
\newcommand{\ToneRealTensor}[1]{\includegraphics[height=.195\textwidth,width=.195\textwidth]{figures/patches_dti/tensor_real_#1.PNG}}
\newcommand{\ToneGenTensor}[1]{\includegraphics[height=.195\textwidth,width=.195\textwidth]{figures/patches_dti/tensor_gen_#1.PNG}}

\usepackage{todonotes}

\begin{document}
\title{Manifold-Aware CycleGAN for High-Resolution Structural-to-DTI Synthesis\thanks{This work was supported financially by the R\'eseau de Bio-Imagerie du Qu\'ebec (RBIQ), the Research Council of Canada (NSERC), the Fonds de Recherche du Qu\'ebec (FQRNT), ETS Montreal, and NVIDIA with the donation of a GPU.}}

\author{Benoit Anctil-Robitaille\inst{1}, Christian Desrosiers\inst{1}, Herve Lombaert\inst{1}}

\authorrunning{B. Anctil-Robitaille et al.}
%
\institute{Department of IT and Software Engineering, ETS Montreal, Canada \email{benoit.anctil-robitaille.1@ens.etsmtl.ca}}
\maketitle

\begin{abstract}
Unpaired image-to-image translation has been applied successfully to natural images but has received very little attention for manifold-valued data such as in diffusion tensor imaging (DTI). The non-Euclidean nature of DTI prevents current generative adversarial networks (GANs) from generating plausible images and has mainly limited their application to diffusion MRI scalar maps, such as fractional anisotropy (FA) or mean diffusivity (MD). Even if these scalar maps are clinically useful, they mostly ignore fiber orientations and therefore have limited applications for analyzing brain fibers. Here, we propose a manifold-aware CycleGAN that learns the generation of high-resolution DTI from unpaired T1w images. We formulate the objective as a Wasserstein distance minimization problem of data distributions on a Riemannian manifold of symmetric positive definite 3$\times$3 matrices SPD(3), using adversarial and cycle-consistency losses. To ensure that the generated diffusion tensors lie on the SPD(3) manifold, we exploit the theoretical properties of the exponential and logarithm maps of the Log-Euclidean metric. We demonstrate that, unlike standard GANs, our method is able to generate realistic high-resolution DTI that can be used to compute diffusion-based metrics and potentially run fiber tractography algorithms. To evaluate our model's performance, we compute the cosine similarity between the generated tensors principal orientation and their ground-truth orientation, the mean squared error (MSE) of their derived FA values and the Log-Euclidean distance between the tensors. We demonstrate that our method produces 2.5 times better FA MSE than a standard CycleGAN and up to 30\% better cosine similarity than a manifold-aware Wasserstein GAN while synthesizing sharp high-resolution DTI.

\end{abstract}
\section{Introduction}

Unpaired image-to-image translation and image synthesis have been widely used in medical imaging \cite{Yi2019}. Whether they are employed to generate missing modalities, normalize images or enhance images quality and resolution, generative adversarial networks (GANs) \cite{Goodfellow2014} have been proven effective in multiple challenging medical image analysis tasks. However, they have been mainly studied on real-valued images, thus impeding the development of applications for manifold-valued data\index{manifold-valued data} such as diffusion tensor images (DTI). Despite the growing interest in the brain's structural connectivity, applications of GANs to DTI have been mostly limited to generating derived scalar maps like fractional anisotropy (FA) and mean diffusivity (MD), which ignore the fibers' orientation and provide limited insights on their structural organization. 

Among the literature, \cite{Gu2019} investigates the generation of diffusion MRI scalar maps from T1w images using a CycleGAN \cite{Zhu2017}. The authors show that the structural and diffusion spaces share a sufficient amount of information to be able to synthesize realistic FA and MD maps from downsampled T1w images. In \cite{Zhong2020}, dual GANs and Markovian discriminators are used to harmonize multi-site FA and MD maps of neonatal brains. They demonstrate that using a GAN-like architecture can better capture the complex non-linear relations between multiple domains than standard normalization methods. 

While the previously mentioned works present applications of GANs on DTI-derived metrics, they do not tackle the challenge of generating DTI. Being able to synthesize such images would unlock a vast amount of useful methods that are already well studied on real-valued modalities, while preserving all the geometrical information encoded in the diffusion tensors. However, DTI data is manifold-valued: the data of each voxel lies on a Riemannian manifold of symmetric positive definite 3$\times$3 matrices, i.e., the SPD(3) manifold. The non-Euclidean nature of DTI prevents standard GANs from generating plausible images as there is no guarantee that the generated diffusion tensors lie on the SPD(3) manifold. A solution presented in \cite{Arsigny2006} is to employ the Log-Euclidean metric to accurately process data on the SPD(n) manifold. By using the $\log$ and $\exp$ projections proposed in \cite{Arsigny2006}, one can apply Euclidean operations on tensors and guarantee that the resulting tensors will lie on such manifold. Those computationally efficient mapping operations form an interesting framework for manifold-valued data learning, and have been used in \cite{Huang2016} to develop a deep neural network called SPDNet which learns discriminative SPD matrices. With the help of the matrix backpropagation of spectral layers defined in \cite{Ionescu}, they designed a network that learns data on SPD(n). Nonetheless, SPDNet \cite{Huang2016} is limited to single SPD matrix learning and cannot help in learning multiple spatially-organized SPD matrices as it is the case with DTI.

Related to our work, \cite{Huang2019} proposes a manifold-aware Wasserstein GAN for manifold-valued data generation, which leverages the aforementioned $\log$ and $\exp$ mappings. In their work, they generate plausible slices of DTI from noise vectors. By comparing the produced images of their network with those produced by a regular GAN, one can clearly see that the manifold mappings are necessary to produce credible diffusion tensors. However, the proposed manifold-aware GAN could not provide any additional clinical insights, nor help in understanding the brain's connectivity as the generated images are not conditioned by any real contextual information such as T1w images. Furthermore, \cite{Huang2019} only focuses on the generation of 2D DTIs, which is of limited application for the assessment of the structural organization of the brain's fibers.

This paper presents a novel manifold-aware Wasserstein CycleGAN that generates high-resolution (HR) DTI\index{high-resolution DTI} from unpaired T1w images. Our method leverages the detailed structural information provided by T1w images while constraining the synthesized diffusion tensors\index{diffusion tensors} to lie on the SPD(3) manifold using the Log-Euclidean metric\index{Log-Euclidean Metric}. Specifically, the contributions of this work are as follows:
\begin{itemize}\setlength\itemsep{3pt}
\item We present the first CycleGAN model for the unpaired mapping between images and SPD(3) manifold-valued data.
\item This is also the first deep learning model to generate DTI data from structural MRI. As mentioned before, previous approaches have focused on generating diffusion scalar maps like FA or MD, and not diffusion tensors as in this work.
\end{itemize}
Our proposed manifold-aware CycleGAN method is presented in the next section.

\section{Method}

\begin{figure}[t!]
\centering
\includegraphics[width=.925\textwidth]{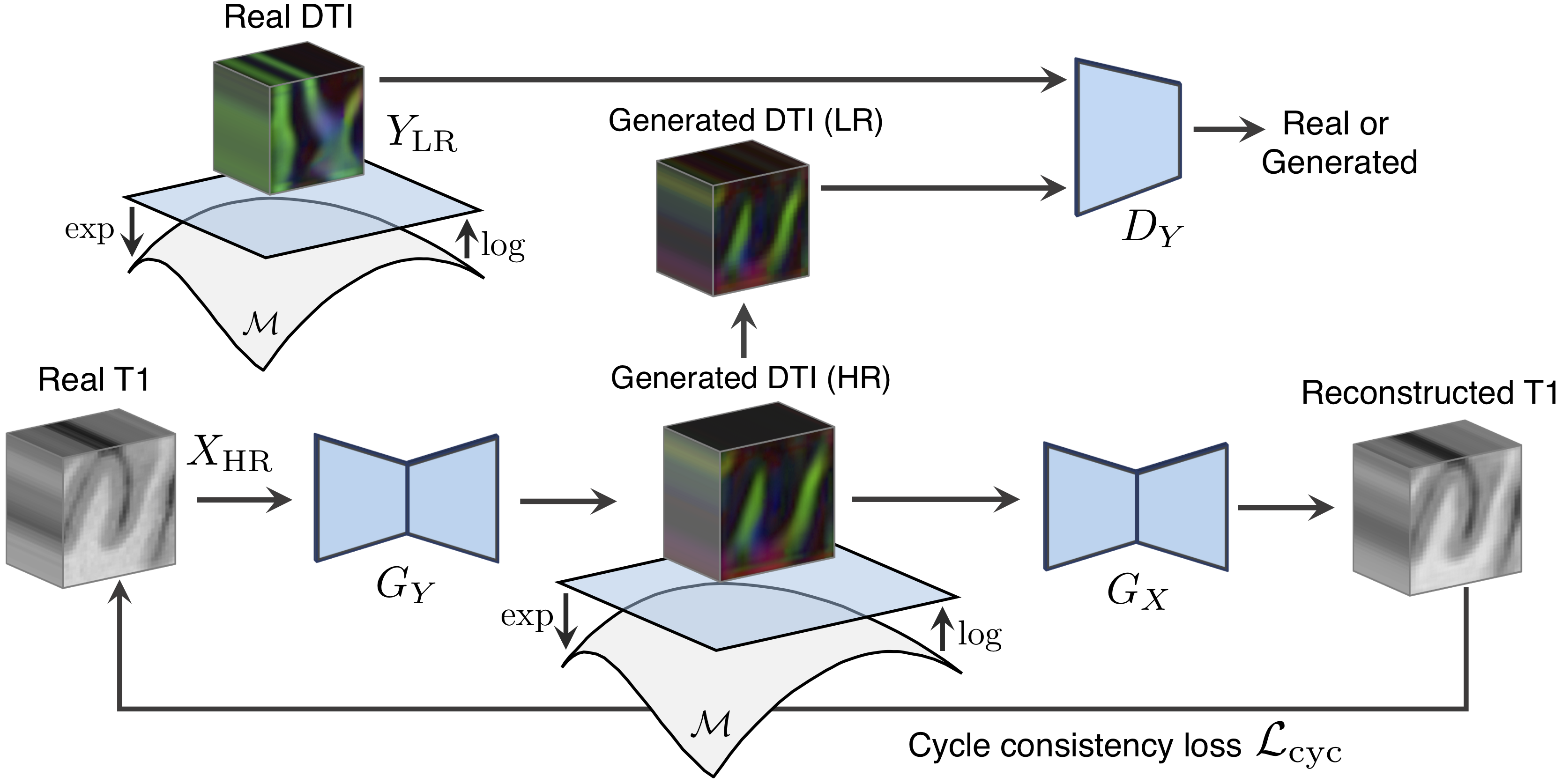}
\caption{The forward cycle of our manifold-aware CycleGAN: $G_Y$ generates high-resolution DTIs on the SPD(3) manifold by projecting its prediction using the $\expId$ and $\logId$ mappings consecutively. $D_Y$ assesses the downsampled generated images and provides adversarial feedback to $G_Y$. $G_X$ tries to reconstruct the original T1w images from 
$G_Y(\xx)$ and supplies high-resolution gradient information to $G_Y$.} \label{architecture}
\end{figure}

Let $\XHR$ be the domain of high-resolution structural images and $\YHR$ be the domain of high-resolution diffusion tensor images. Our goal is to learn mapping functions $\genY: \XHR \mapsto \YHR$ and $\genX: \YHR \mapsto \XHR$ that translate the real-valued domain $\XHR$ into the manifold-valued domain $\YHR$ and the other way around. However, as it is often the case, we do not have access to high-resolution DTI. Thus, we train our model with unpaired training samples $\{\xx_{i}\}^{N}_{i=1}$ where $\xx_{i} \in \XHR $ is a 3D structural image (e.g., T1w), and $\{\yy_{j}\}^{M}_{j=1}$ where $\yy_{j} \in \YLR$ is a DT image with lower resolution. We employ the $\logId$ and $\expId$ mapping to project the generated and the real DTI on the tangent plane at the 3$\times$3 identity matrix, to ensure that $\genY(\xx)$ lies on the SPD(3) manifold and to compare the manifold-valued data distributions as in \cite{Huang2019}. Two discriminators, $\disX$ and $\disY$, assess the quality of the generated images $\genX(\yy)$ and downsampled generated DTI 
$\downarrow\!\genY(\xx)$
with respect to their real data distributions $\genX(\yy)\sim{\Prob_{\XHR}}$ and 
$\downarrow\!\!\genY(\xx) \sim{\Prob_{\logId(\YLR)}}$.
We formulate the objective as a Wasserstein distance minimization problem on the SPD(3) manifold with adversarial and cycle-consistent losses. The adversarial portion of the objective helps $\genX$ and $\genY$ generate images that match the target distribution. On the other hand, the cycle-consistent losses provide high-resolution gradients that are necessary to generate high-resolution DTI with a proper structure. 

\begin{figure}[t!]
\setlength{\tabcolsep}{.5pt}
\begin{large}
\begin{tabular}{cccc}
\ToneReal{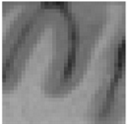} & \ToneReal{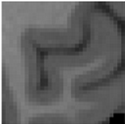} & \ToneReal{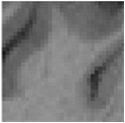} & \ToneReal{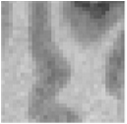} \\
\ToneRec{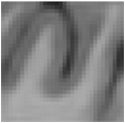} & \ToneRec{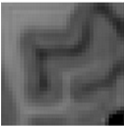} & \ToneRec{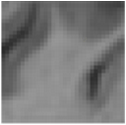} & \ToneRec{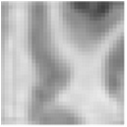}
\end{tabular}
\end{large}
\caption{(\textbf{top row}) Real high-resolution T1w patches and (\textbf{bottom row}) recovered T1w patches.} 
\label{fig:visualization}
\end{figure}

\subsection{Log-Euclidean Metric}

Diffusion tensor matrices are well defined in the Log-Euclidean metric, where a matrix logarithm and exponential can be conveniently processed in one metric and always be mapped back to valid symmetric diffusion tensors \cite{Arsigny2006}. Let $\matr{M} = \UU \Eig \tr{\UU}$ be the eigendecomposition of a symmetric matrix $\matr{M}$. The computation of the logarithm and the exponential of a tensor noted as $\logId$ and $\expId$ are defined as follows:
\begin{align}
    \forall \PP \in S_{++}^*, \ \logId(\PP) & \ = \ \UU\, \log(\Eig)\,\tr{\UU} \, \in \, \Tid\label{log_of_tensor} \\
    \forall \Ss \, \in \, \Tid, \ \expId(\Ss) & \ = \ \UU\,\exp(\Eig)\,\tr{\UU} \, \in \,  S_{++}^*\label{exp_of_tensor}
\end{align}
We use these maps throughout our work to project the generated and real DTI on the SPD(3) manifold and on the tangent plane at the 3$\times$3 identity matrix $\Tid$. Moreover, the Log-Euclidean distance between two tensors $\PP_1$ and $\PP_2$ is defined as:
\begin{align}
   dist(\PP_1, \PP_2) \, = \, \big\|\logId(\PP_1)-\logId(\PP_2)\big\|_2\label{eq:log_distance}.
\end{align}
In our framework, we use this distance to measure the similarity between predicted and real DTI in the tangent plane.   

\subsection{Adversarial Loss}

In a traditional GAN setup \cite{Goodfellow2014}, a generator $G$ and a discriminator $D$ compete in a minimax game where $G$ tries to generate data close to a true data distribution so that $D$ cannot identify if the generated data is real or not. In \cite{Arjovsky2017}, $D$ is replaced by a discriminator that leverages the Wasserstein distance to estimate the similarity between the real and generated data distributions. The Wasserstein GAN (WGAN) architecture tends to stabilize the training as the Wasserstein distance never saturates, and thus always provides relevant gradients to $G$. The adversarial part of our objective follows the WGAN framework and is divided in two separate loss terms, $\LwganX$ and $\LwganY$, respectively for structural images and DTI: 
\begin{align}\label{adversarial_loss}
    & \LwganX(\genX, D_X, \YLR, \XHR) \nonumber\\
    & \qquad \ = \ 
        \EE_{\xx\sim\Prob_{\XHR}}\big[D_X(\xx)\big] \, - \, \EE_{\yy\sim\Prob_{\YLR}} \big[D_X\big(\genX(\uparrow\!\logId(\yy))\big)\big] \\
    & \LwganY(\genY, D_Y, \XHR, \YLR) \nonumber\\
    & \qquad \ = \ 
        \EE_{\yy\sim\Prob_{\YLR}}\big[D_Y(\logId(\yy))\big] \, - \, \EE_{\xx\sim\Prob_{\XHR}} \big[D_Y\big(\!\downarrow\!\genY(\xx)\big)\big]    
\end{align}
where $\uparrow\!$ and $\!\downarrow\!$ indicates trilinear up and downsampling.

In $\LwganX$, $\genX$ generates 3D structural images from high-resolution DTI projected on the identity-based tangent plane using the $\logId$ mapping. Since we only have samples from the low-resolution data distribution $\yy\sim\Prob_{\YLR}$, the real DTI data is upsampled via trilinear interpolation before being fed to $\genX$. In the same loss, $D_X$ measures the Wasserstein distance between the data distribution of generated and real structural images. Likewise, in $\LwganY$, $D_Y$ computes the Wasserstein distance between the distribution of downsampled generated DTI and real low-resolution DTI in the tangent plane, using the Log-Euclidean metric \cite{Arsigny2006}. 

\subsection{Cycle Consistency Loss}

The adversarial loss alone is not sufficient to drive the generation of high-resolution DTIs. Indeed, the discriminator $D_Y$ only assesses downsampled DT images so its feedback cannot help $\genY$ improve beyond a certain level of detail. To mitigate this problem, we introduce a second loss that enforces the forward and backward cycle consistency of the network \cite{Zhu2017}, and provides high-resolution gradient information. In our case, the forward cycle consistency ensures that, from the translated DTI 
$\genY(\xx)$,
we are able to reconstruct the corresponding original structural images $\xx\sim\Prob_{\XHR}$ which are originally in high resolution. The backward cycle consistency ensures that we are able to reconstruct the original upsampled DTI $\yy\sim{\Prob_{\logId(\YLR)}}$ from the translated structural images $\genX(\uparrow\!\logId(\yy))$. The total cycle consistency loss is defined as
\begin{align}\label{lossCyc}
    & \Lcyc(\genY, \genX) \ = \ \lambda_{\mr{cyc}_{X}} \EE_{\xx \sim \Prob_{\XHR}} \big[\norm{\genX(\genY(\xx)) \, - \, \xx}_1\big] \nonumber\\    
   & \qquad \quad \ + \ \lambda_{\mr{cyc}_{Y}} \EE_{\yy \sim \Prob_{\YLR}} 
    \big[\norm{\genY(\genX(\uparrow\!\logId(\yy))\big) \, - \, \uparrow\!\logId(\yy)}_1\big]
\end{align}
Here, $\lambda_{\mr{cyc}_{X}}$ and $\lambda_{\mr{cyc}_{Y}}$ balance the contribution of the forward and backward cycles respectively and have been empirically set to a value of 3 and 1. Note that $\ell_1$ norm is employed instead of $\ell_2$ to make the loss less sensitive to large reconstruction errors.

\begin{figure}[t!]
\includegraphics[width=\textwidth]{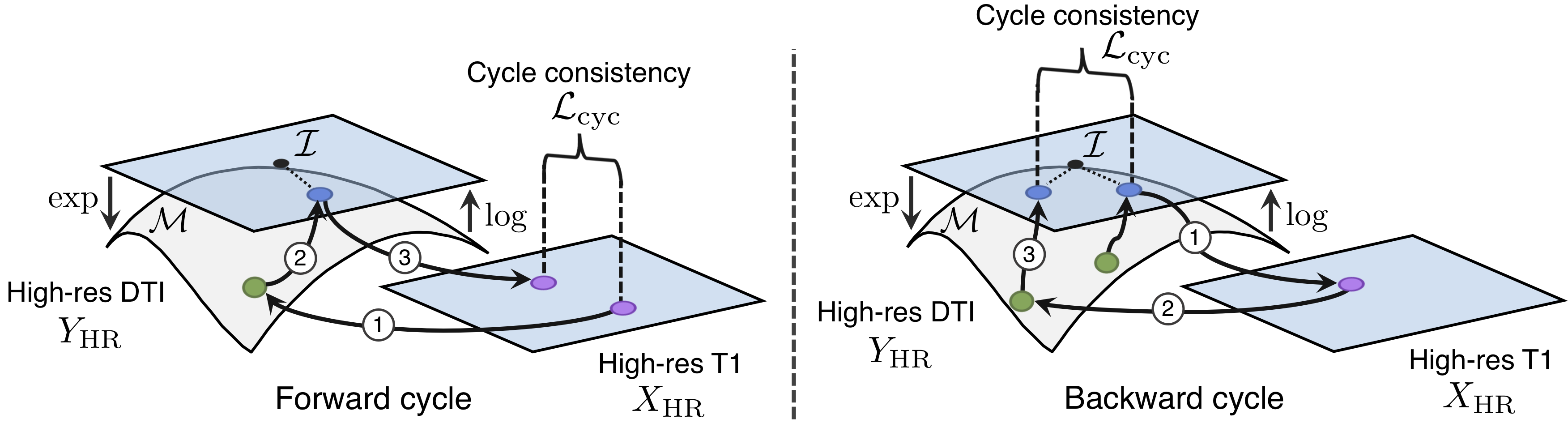}
\caption{On the left, our forward cycle: (1) A T1w image is translated into a high-resolution DTI where each voxel belongs to the SPD(3) manifold, (2) the tensors are projected to $\Tid$ using the $\logId$ map and (3) the image is translated back to the T1w domain where $\Lcyc$ is computed. On the right, our backward cycle: (1) An upsampled DTI on $\Tid$ is translated to the T1w domain, (2) the generated T1w image is translated back to DTI and (3) the recovered DTI is projected to $\Tid$ where $\Lcyc$ is computed.} \label{manifold_cycle_consistency}
\end{figure}

\subsection{Manifold-Aware Wasserstein CycleGAN}
Our full objective is
\begin{equation}\label{loss}
\begin{split}
   & \Loss(\genX, \genY, \disX, \disY) \ = \ \lambda_{\mr{GAN}_{X}}\LwganX(\genX, D_X, \YLR, \XHR) \\ 
    & \qquad\qquad \ + \ \lambda_{\mr{GAN}_{Y}}\LwganY(\genY, D_Y, \XHR, \YLR) 
        \ + \  \Lcyc(\genY, \genX)
 \end{split}
\end{equation}
The adversarial parts $\Lwgan(\genX, D_X, \YLR, \XHR)$ and $\Lwgan(\genY, D_Y, \XHR, \YLR)$  of our full objective guide the generators $\genX$ and $\genY$ towards the synthesis of images close to their real data distributions using a Wasserstein distance on the SPD(3) manifold. The cycle consistency denoted as $\Lcyc(\genY, \genX)$ gives fine-grained retro-action that helps generate HR DTI while preventing mode collapse. Once training is done, 
the high-resolution DTI $\yy_{\mathrm{HR}}$ of a structural image $\xx$ can be obtained by applying the exponential map to the DTI generator output: $\yy_{\mathrm{HR}} = \expId(\genY(\xx))$.

\section{Experiments}
\mypar{Data} We employ the pre-processed T1-weighted (T1w) and diffusion MRI (dMRI) data of 1,065 patients from the HCP1200 release of the Human Connectome Project \cite{VanEssen2013} to evaluate our manifold-aware CycleGAN. The T1w (0.7 mm isotropic, FOV = 224mm, matrix = 320, 256 sagittal slices in a single slab) and diffusion (sequence = Spin-echo EPI, repetition time (TR) = 5520 ms, echo time (TE) = 89.5ms, resolution = 1.25$\times$1.25$\times$1.25 mm$^3$ voxels) data acquisition was done using a Siemens Skyra 3T scanner \cite{Sotiropoulos2013} and pre-processed following \cite{Glasser2013}. The diffusion tensors were fitted using DSI Studio toolbox \cite{Jiang2005}. Both T1w and DTI were decomposed in overlapping patches of $32^3$ voxels centered on the foreground.

\mypar{Experiments Setup} We used 50,000 unpaired T1w and DTI patches randomly selected among 1,055 subjects as our training set. For the validation set and the test set, we took paired T1w and DTI patches covering the full brain of respectively 3 and 2 randomly chosen subjects. We compared our manifold-aware CycleGAN (MA-CycleGAN) method with two baselines: a manifold-aware Wasserstein GAN without cycle (MA-GAN) and a Wasserstein CycleGAN without the $\logId$ and $\expId$ mappings. These baseline methods allow us to assess the impact of both the cycle consistency and the manifold mapping. We measure the quality of the generated HR DTI by computing three metrics: 1) the mean cosine similarity between the principal eigenvectors of the generated images and their ground-truth, 2) the mean squared error between the FA of the generated images and their ground-truth, and 3) the mean Log-Euclidean distance between the generated tensors and their ground-truth following Equation~(\ref{eq:log_distance}). Because the principal eigenvector's direction is more relevant at voxels with higher anisotropy, cosine similarity is measured at three different FA thresholds taken on the ground-truth images: FA $\geq$ 0 (all voxels), only voxels with FA $\geq$ 0.2, and only voxels with FA $\geq$ 0.5. An FA threshold near 0.2 is commonly used for tract-based analysis of white matter \cite{taoka2009fractional}.

\begin{figure}[t!]
\setlength{\tabcolsep}{0.5pt}
\renewcommand{\arraystretch}{0.5}
\begin{large}
\begin{tabular}{ccccc}
\ToneRealFA{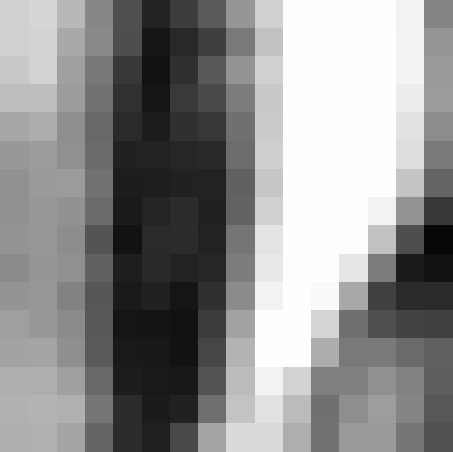} & \ToneRealFA{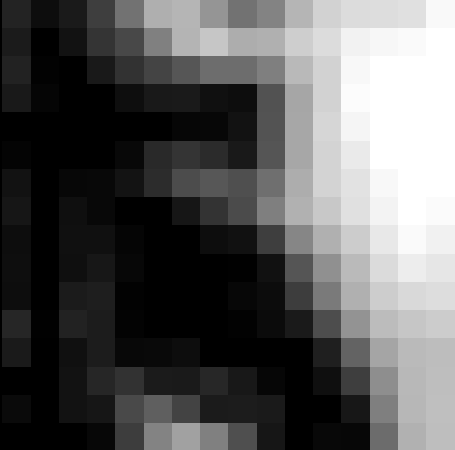} & \ToneRealFA{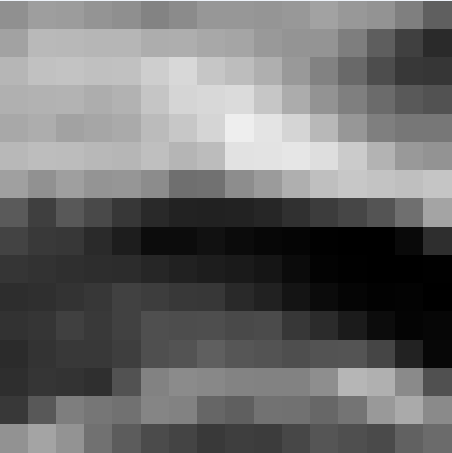} & \ToneRealFA{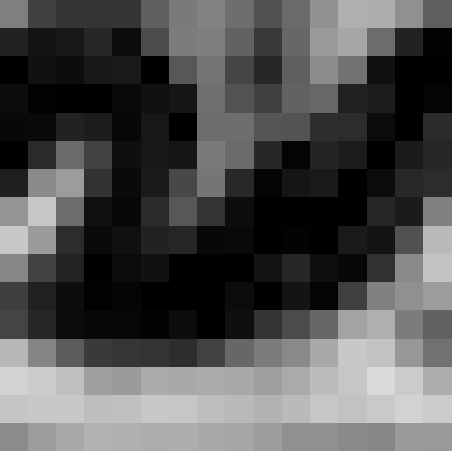} & \ToneRealFA{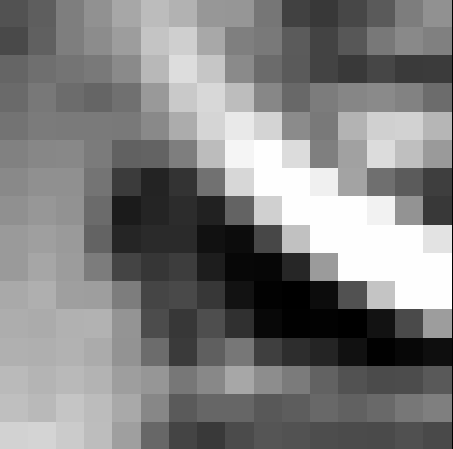}\\
\ToneGenFA{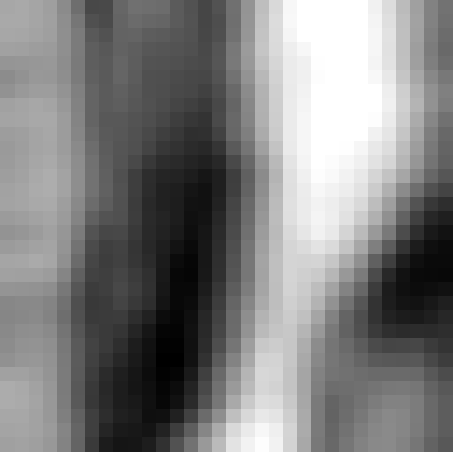} & \ToneGenFA{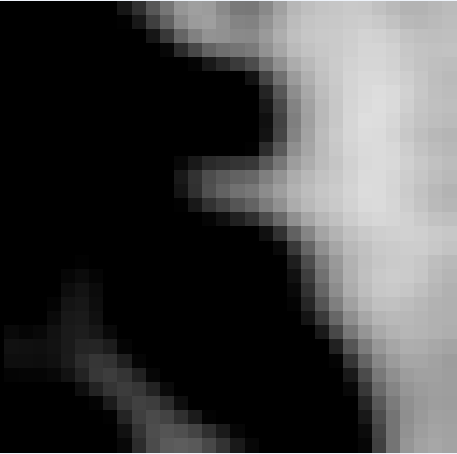} & \ToneGenFA{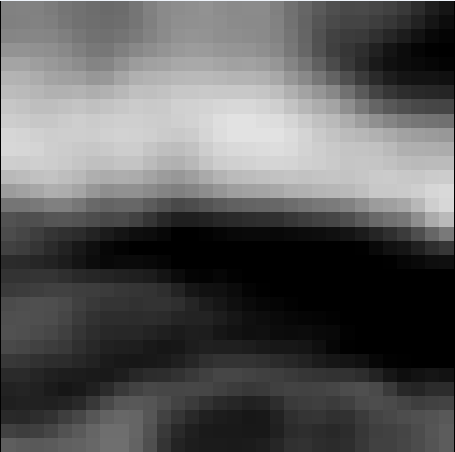} & \ToneGenFA{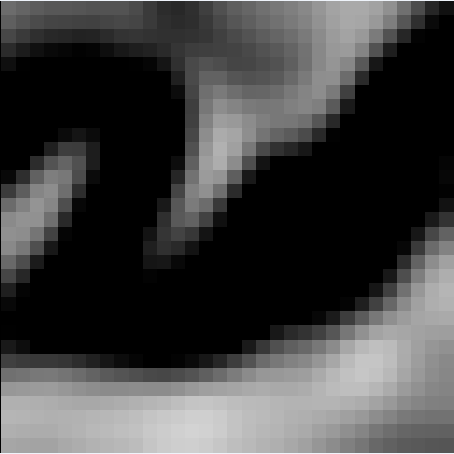} & \ToneGenFA{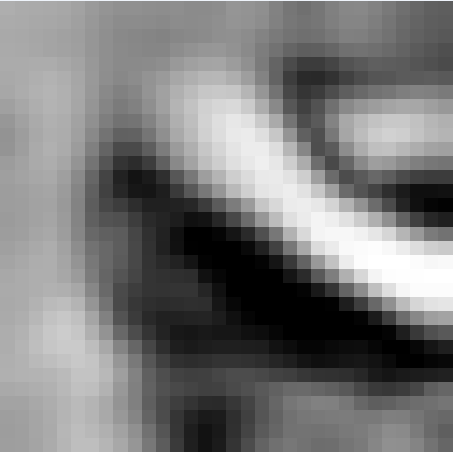}\\
\ToneRealTensor{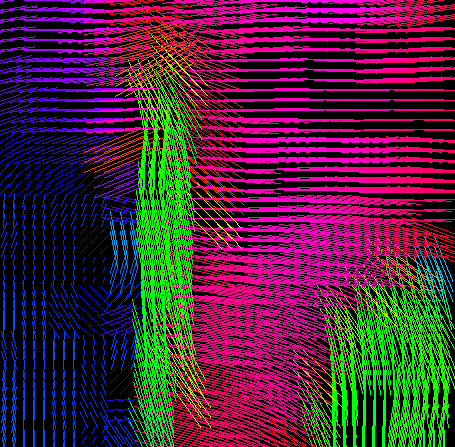} & \ToneRealTensor{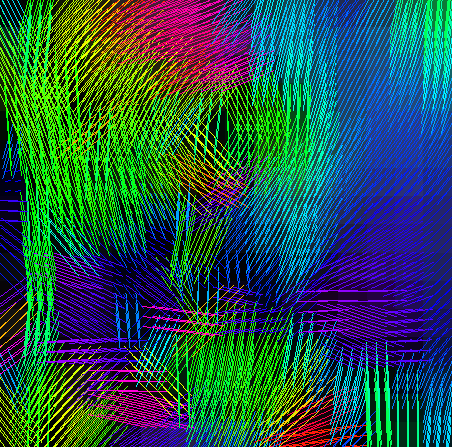} & \ToneRealTensor{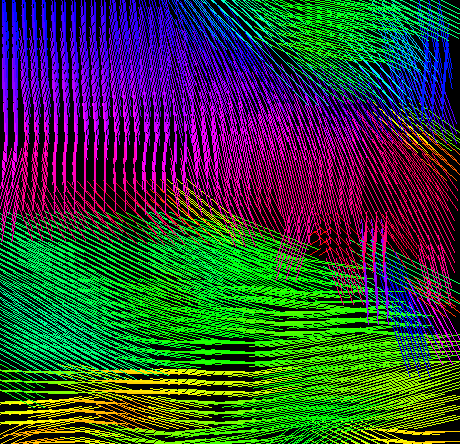} & \ToneRealTensor{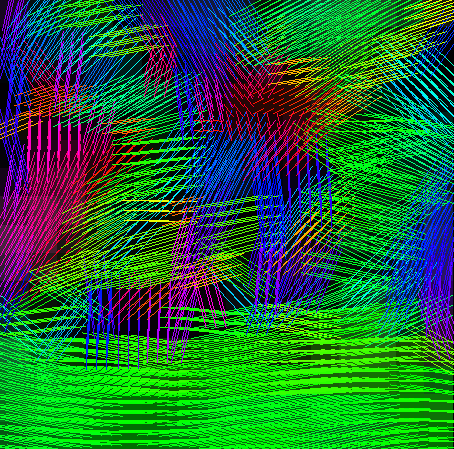} & \ToneRealTensor{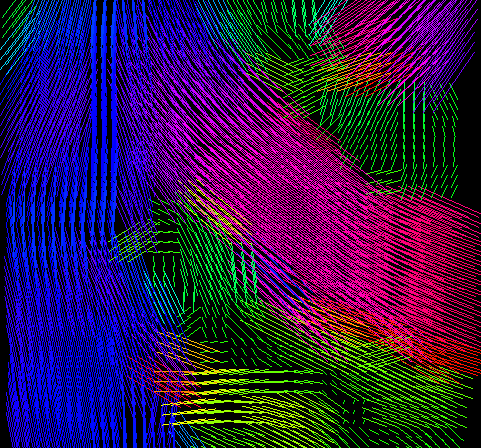}\\
\ToneGenTensor{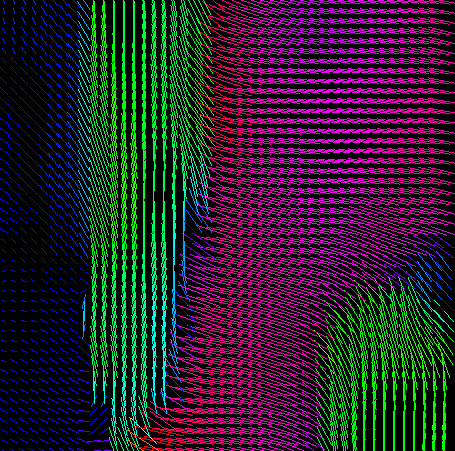} & \ToneGenTensor{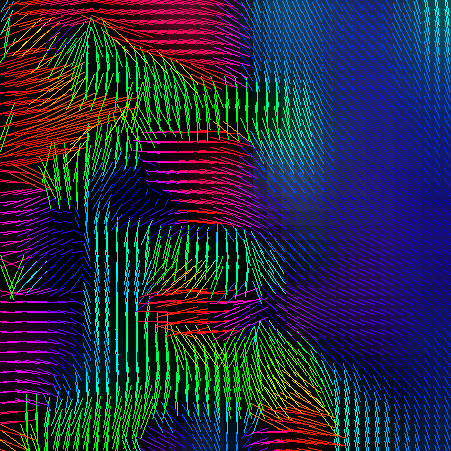} & \ToneGenTensor{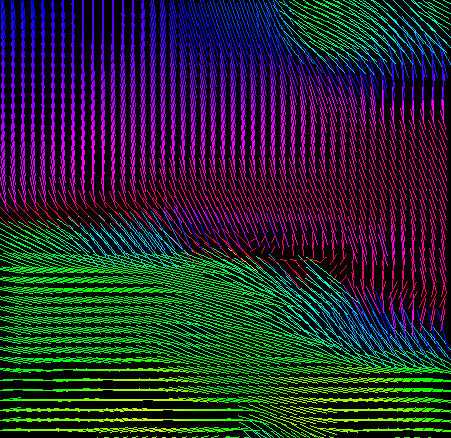} & \ToneGenTensor{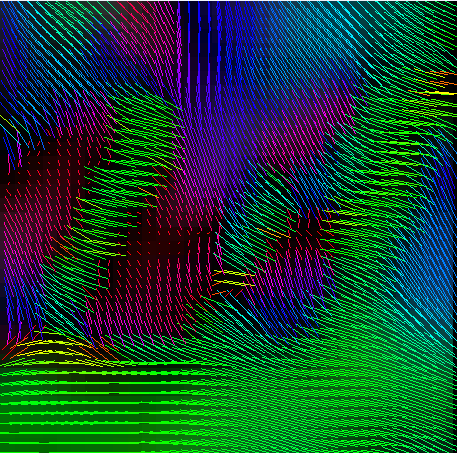} & \ToneGenTensor{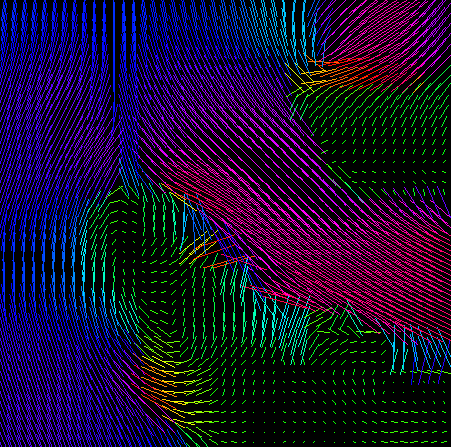}
\end{tabular}
\end{large}
\caption{(\textbf{top row}) Real low-resolution FA, (\textbf{second row}) generated high-resolution FA, (\textbf{third row}) real low-resolution color orientation tensors, and (\textbf{bottom row}) generated high-resolution color orientation tensors. Best viewed in color.} \label{fig:tensor_fa}
\end{figure}

While the cosine similarity allows us to evaluate the precision of the predicted orientation of the generated tensors, the mean squared error on the derived FA highlights the network's ability to estimate local diffusion anisotropy. As for the Log-Euclidean distance, it takes into account both the orientation and the anisotropy of the tensors. Furthermore, a qualitative inspection of the generated tensors and FA is performed in Figure \ref{fig:tensor_fa}.

\mypar{Implementation Details} Both generators are based on the Unet implementation from \cite{Ronneberger2015} where the convolutions have been changed to 3D convolutions. In addition, we changed the last activation layers to fit the scale of the generated data and the number of channels with respect to our inputs and outputs. For $\genX$, we use a sigmoid as the final activation function to generate values in the range [0,1]. In the case of $\genY$, we use a hard hyperbolic tangent activation function. 
Furthermore, to guarantee that the synthesized tensors can be decomposed following Equations (\ref{log_of_tensor}) and (\ref{exp_of_tensor}), we convert the 9-channels output of $\genY$ into a 3$\times$3 matrix $\matr{Y}$ and make this matrix symmetric as follows: $\matr{Y}' = \frac{1}{2}(\matr{Y} + \tr{\matr{Y}})$. For our discriminators, we employ a Resnet-18 architecture \cite{He2016} where all convolutions have been changed to 3D convolutions. Both the generators and discriminators were trained for 30 epochs with the Adam optimizer \cite{Kingma2015} and a batch size of 8. A starting learning rate of $1\times10^{-4}$ was used jointly with a reduce-on-plateau strategy. To stabilize the training of our network, we pre-trained the generators independently with 25,000 paired patches randomly sampled from 5 subjects during 10 epochs. The paired patches have been computed from aligned high-resolution structural T1w images and upsampled DTIs. The alignment was performed using FLIRT \cite{Jenkinson2002,Jenkinson2001} from FSL \cite{Jenkinson2012}.

\begin{table*}[t]
\caption{Fractional anisotropy mean squared error (FA MSE), Log-Euclidean distance, and cosine similarity between principal eigenvectors of compared methods: manifold-aware Wasserstein GAN (MA-GAN), Wasserstein CycleGAN without the $\logId$ and $\expId$ mappings (CycleGAN), and our manifold-aware Wasserstein CycleGAN (MA-CycleGAN). A smaller FA MSE and Log-Euclidean distance corresponds to a superior performance while a higher cosine similarity is better. Since the principal eigenvector's direction is more relevant at voxels with higher diffusivity, we report cosine similarity at increasing FA thresholds of 0, 0.2 and 0.5.}\label{results_tab}
\centering
\begin{footnotesize}
\renewcommand{\arraystretch}{1.2}
\begin{tabular}[t]{p{3.8cm}C{1.3cm}C{1.7cm}C{1.3cm}C{1.5cm}C{1.5cm}}
\toprule
&  &  & \multicolumn{3}{c}{\textbf{Cosine Similarity}}\\
\cmidrule(lr){4-6}
\textbf{Models} & \textbf{FA MSE} & \textbf{Log Distance} & FA $\geq$ 0 & FA $\geq$ 0.2 & FA $\geq$ 0.5 \\
\midrule
MA-GAN & 0.0277 & 0.6067 & 0.4512  & 0.6369 & 0.6700\\
CycleGAN &  0.0431 & 0.5699 & 0.5371 & 0.6717 & 0.7064\\
\textbf{MA-CycleGAN (ours)} & \textbf{0.0172} & \textbf{0.5515} & \textbf{0.5846} & \textbf{0.7217} & \textbf{0.8041}\\
\bottomrule
\end{tabular}
\end{footnotesize}
\end{table*}

\begin{figure}[ht!]
\centering
\includegraphics[width=.99\textwidth]{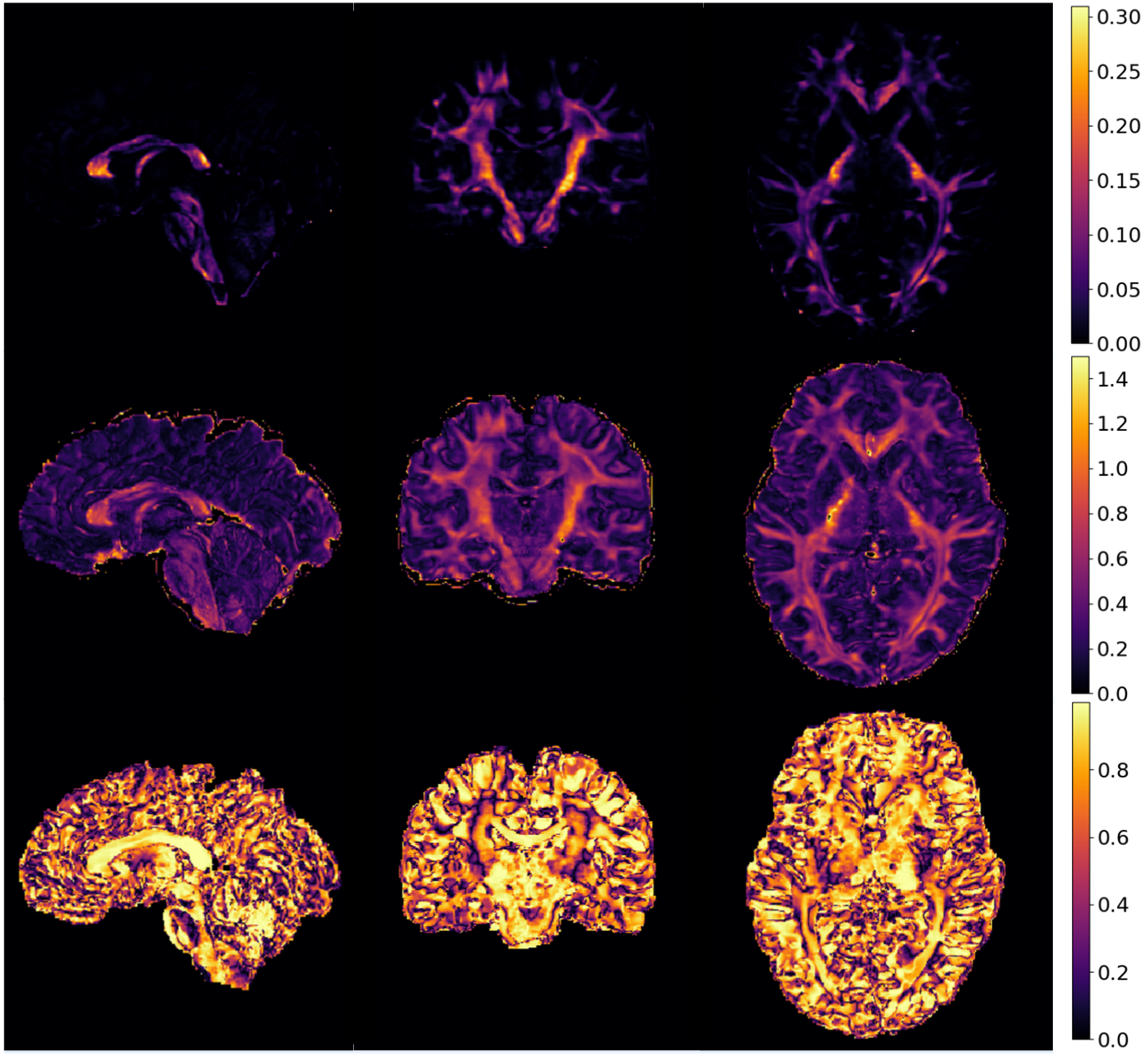}
\caption{Metrics on the sagittal, coronal and axial slices between the generated HR DTI of a random evaluation subject and its interpolated ground-truth. (\textbf{top row}) FA MSE, (\textbf{middle row}) Log-Euclidean distance and (\textbf{bottom row}) cosine similarity.} \label{fig:metrics}
\end{figure}

\mypar{Fibers Orientation Analysis} To evaluate the predicted tensor orientation, we compute the cosine similarity between the principal orientation of each generated tensor and its ground-truth, for FA threshold 0.0, 0.2 and 0.5. Results in Table~\ref{results_tab} show that our method performs better than the two baselines, yielding average improvements in cosine similarity (for FA $\geq$ 0.0, 0.2, 0.5) of 0.133, 0.085, 0.134 compared to the manifold-aware GAN without cycle and 0.048, 0.050, 0.100 compared to the CycleGAN without manifold mapping. As mentioned before, a good estimation of main diffusion orientation is generally more important at voxels with greater diffusivity. Hence, the cosine similarity for FA $\geq$ 0.2 and FA $\geq$ 0.5 is a better indicator of performance than for FA $\geq$ 0.0. We see in Table~\ref{results_tab} that our model's estimation of fiber orientation improves with a higher FA threshold, reaching a similarity of 0.804 for voxels with FA $\geq$ 0.5. This can be observed in Figure \ref{fig:metrics}, which gives the FA MSE, Log-Euclidean distance and cosine similarity at each voxel of sagittal, coronal and axial slices from the same subject. As can be seen, orientations in regions with typically high FA, like the corpus callosum, are better predicted by our model than those in regions with lower FA.

\begin{figure}[t!]
\includegraphics[width=\textwidth]{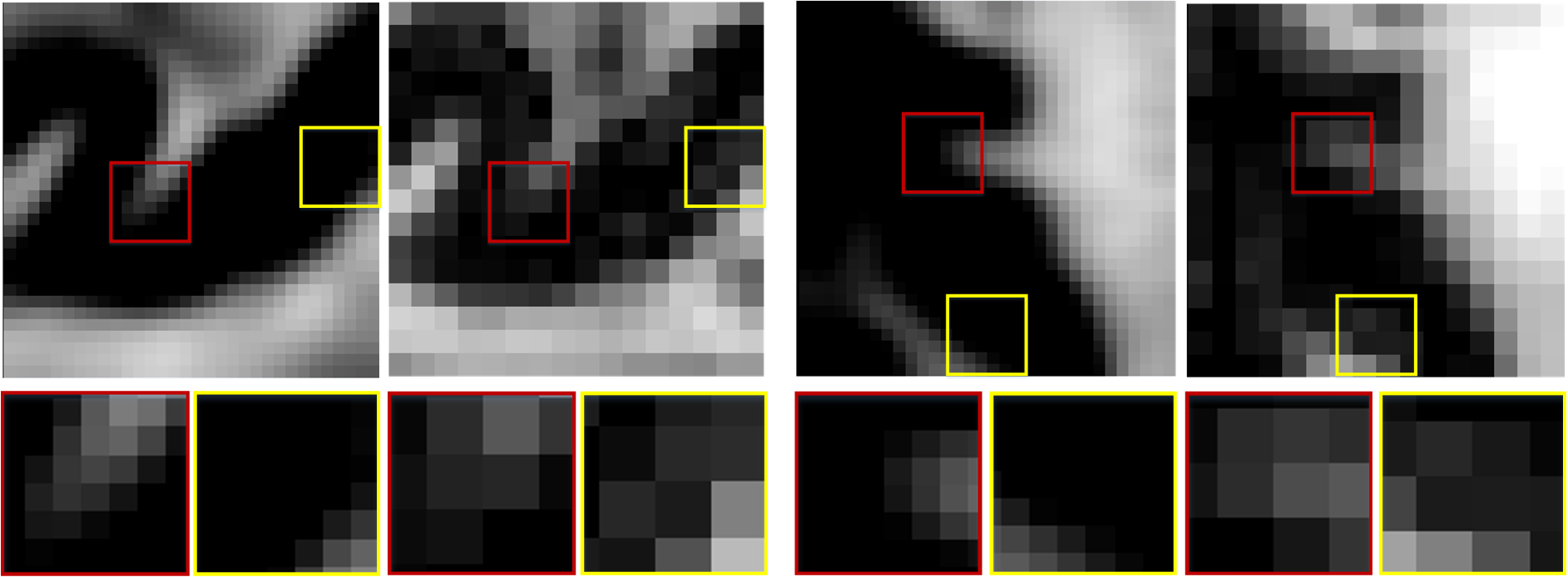}
\caption{Pairs of derived high-resolution FA and their low-resolution ground-truth. The HR FA shows sharper edges and lower partial volume effect.} \label{fig:pve_zoom}
\end{figure}

\mypar{FA Analysis} Next, we evaluate the fractional anisotropy (FA) of generated HR DTI. FA is one of the most commonly used DT-derived metrics, thus an accurate prediction of this metric is critical. Table \ref{results_tab} shows the mean squared error (FA MSE) obtained by our method and the two baselines. Once more, our manifold-aware CycleGAN outperforms the manifold-aware GAN and the standard CycleGAN with an MSE of 0.172 compared to 0.0277 and 0.0431. Furthermore, we observe that the two methods using manifold mapping perform better than the standard CycleGAN. This performance gap is due to the fact that, without projecting the generated data on the SPD(3) manifold using the Log-Euclidean metric, there is no guarantee that the generated tensors lie on such manifold. Consequently, the tensors eigenvalues used in the computation of FA are not strictly positive, which increases the differences between the generated FA and the ground-truth. In addition, as seen in Figures \ref{fig:tensor_fa} and \ref{fig:pve_zoom}, generating high-resolution DTI helps reducing partial volume effect which is known to impact subsequent analysis \cite{pfefferbaum2003increased}. However, as shown in Figure \ref{fig:metrics}, our model tends to underestimate the FA in white matter regions where the FA is further away from the mean value, despite the good estimation of fiber orientation.

\section{Discussion and Conclusion}

In this paper we proposed a novel manifold-aware CycleGAN that successfully leverages the Log-Euclidean metric and the structural information of T1w images to generate realistic high-resolution DTI. Our method outperformed the manifold-aware GAN and the standard CycleGAN architecture in terms of tensor principal orientation estimation, Log-Euclidean distance and MSE of derived FA. These results not only confirm that projecting the generated DTI on the SPD(3) manifold helps producing plausible diffusion tensors but also that the extra structural information provided by the T1w data is necessary to synthesize high-resolution DTI. Although physiological evidence remains limited, it is shown in \cite{schyboll2018impact} that fiber orientations have an impact on T1w image intensities. Our results further suggest that T1w images may contain information on the high-level geometry of fiber tracts, which can be learned by the network to estimate the diffusion properties and orientation. However, a deeper investigation is required to validate this hypothesis.

We believe that our method is an important contribution to medical image computing as it unlocks a vast number of applications on manifold-valued data. As future work, we plan on extending our method to other Riemannian manifolds such as the statistical manifold for orientation distribution function estimation. Furthermore we will investigate the integrity of our generated data with common downstream tasks such as tractography and fiber bundles segmentation.

%
%
%
%
\bibliographystyle{splncs04}
\bibliography{references}

\begin{thebibliography}{10}
\providecommand{\url}[1]{\texttt{#1}}
\providecommand{\urlprefix}{URL }
\providecommand{\doi}[1]{https://doi.org/#1}

\bibitem{Arjovsky2017}
Arjovsky, M., Chintala, S., Bottou, L.: {Wasserstein GAN}  (2017)

\bibitem{Arsigny2006}
Arsigny, V., Fillard, P., Pennec, X., Ayache, N.: {Log-Euclidean metrics for
  fast and simple calculus on diffusion tensors}. Magnetic Resonance in
  Medicine  (2006)

\bibitem{Glasser2013}
Glasser, M.F., Sotiropoulos, S.N., Wilson, J.A., Coalson, T.S., Fischl, B.,
  Andersson, J.L., Xu, J., Jbabdi, S., Webster, M., Polimeni, J.R., {Van
  Essen}, D.C., Jenkinson, M.: {The minimal preprocessing pipelines for the
  Human Connectome Project}. NeuroImage  (2013)

\bibitem{Goodfellow2014}
Goodfellow, I.J., Pouget-Abadie, J., Mirza, M., Xu, B., Warde-Farley, D.,
  Ozair, S., Courville, A., Bengio, Y.: {Generative adversarial nets}. In:
  Advances in Neural Information Processing Systems. Neural information
  processing systems foundation (2014)

\bibitem{Gu2019}
Gu, X., Knutsson, H., Nilsson, M., Eklund, A.: {Generating Diffusion MRI Scalar
  Maps from T1 Weighted Images Using Generative Adversarial Networks}. Tech.
  rep. (2019)

\bibitem{He2016}
He, K., Zhang, X., Ren, S., Sun, J.: {Deep residual learning for image
  recognition}. In: Proceedings of the IEEE Computer Society Conference on
  Computer Vision and Pattern Recognition. IEEE Computer Society (2016)

\bibitem{Huang2016}
Huang, Z., {Van Gool}, L.: {A riemannian network for SPD matrix learning}. 31st
  AAAI Conference on Artificial Intelligence, AAAI 2017  (2017)

\bibitem{Huang2019}
Huang, Z., Wu, J., {Van Gool}, L.: {Manifold-Valued Image Generation with
  Wasserstein Generative Adversarial Nets}. Proceedings of the AAAI Conference
  on Artificial Intelligence pp. 3886--3893 (2019)

\bibitem{Ionescu}
Ionescu, C., Vantzos, O., Sminchisescu, C.: {Matrix backpropagation for deep
  networks with structured layers}. Tech. rep. (2015)

\bibitem{Jenkinson2002}
Jenkinson, M., Bannister, P., Brady, M., Smith, S.: {Improved Optimization for
  the Robust and Accurate Linear Registration and Motion Correction of Brain
  Images}. NeuroImage  (2002)

\bibitem{Jenkinson2012}
Jenkinson, M., Beckmann, C.F., Behrens, T.E.J., Woolrich, M.W., Smith, S.M.:
  {Review FSL}. NeuroImage  (2012)

\bibitem{Jenkinson2001}
Jenkinson, M., Smith, S.: {A global optimisation method for robust affine
  registration of brain images}. Medical Image Analysis  (2001)

\bibitem{Jiang2005}
Jiang, H., {Van Zijl}, P.C.M., Kim, J., Pearlson, G.D., Mori, S.: {DtiStudio:
  Resource program for diffusion tensor computation and fiber bundle tracking}
  (2005)

\bibitem{Kingma2015}
Kingma, D.P., Ba, J.L.: {Adam: A method for stochastic optimization}. In: 3rd
  International Conference on Learning Representations, ICLR 2015 - Conference
  Track Proceedings. International Conference on Learning Representations, ICLR
  (2015)

\bibitem{pfefferbaum2003increased}
Pfefferbaum, A., Sullivan, E.V.: Increased brain white matter diffusivity in
  normal adult aging: relationship to anisotropy and partial voluming. Magnetic
  Resonance in Medicine: An Official Journal of the International Society for
  Magnetic Resonance in Medicine  (2003)

\bibitem{Ronneberger2015}
Ronneberger, O., Fischer, P., Brox, T.: {U-net: Convolutional networks for
  biomedical image segmentation}. In: Lecture Notes in Computer Science
  (including subseries Lecture Notes in Artificial Intelligence and Lecture
  Notes in Bioinformatics) (2015)

\bibitem{schyboll2018impact}
Schyboll, F., Jaekel, U., Weber, B., Neeb, H.: The impact of fibre orientation
  on t1-relaxation and apparent tissue water content in white matter. Magnetic
  Resonance Materials in Physics, Biology and Medicine  \textbf{31}(4),
  501--510 (2018)

\bibitem{Sotiropoulos2013}
Sotiropoulos, S.N., Jbabdi, S., Xu, J., Andersson, J.L., Moeller, S., Auerbach,
  E.J., Glasser, M.F., Hernandez, M., Sapiro, G., Jenkinson, M., Feinberg,
  D.A., Yacoub, E., Lenglet, C., {Van Essen}, D.C., Ugurbil, K., Behrens, T.E.:
  {Advances in diffusion MRI acquisition and processing in the Human Connectome
  Project}. NeuroImage  (2013)

\bibitem{taoka2009fractional}
Taoka, T., Morikawa, M., Akashi, T., Miyasaka, T., Nakagawa, H., Kiuchi, K.,
  Kishimoto, T., Kichikawa, K.: Fractional anisotropy--threshold dependence in
  tract-based diffusion tensor analysis: evaluation of the uncinate fasciculus
  in alzheimer disease. American Journal of Neuroradiology  \textbf{30}(9),
  1700--1703 (2009)

\bibitem{VanEssen2013}
{Van Essen}, D.C., Smith, S.M., Barch, D.M., Behrens, T.E., Yacoub, E.,
  Ugurbil, K.: {The WU-Minn Human Connectome Project: An overview}. NeuroImage
  (2013)

\bibitem{Yi2019}
Yi, X., Walia, E., Babyn, P.: {Generative adversarial network in medical
  imaging: A review}. Medical Image Analysis  \textbf{58} (2019)

\bibitem{Zhong2020}
Zhong, J., Wang, Y., Li, J., Xue, X., Liu, S., Wang, M., Gao, X., Wang, Q.,
  Yang, J., Li, X.: {Inter-site harmonization based on dual generative
  adversarial networks for diffusion tensor imaging: Application to neonatal
  white matter development}. BioMedical Engineering Online  \textbf{19} (2020)

\bibitem{Zhu2017}
Zhu, J.Y., Park, T., Isola, P., Efros, A.A.: {Unpaired Image-to-Image
  Translation Using Cycle-Consistent Adversarial Networks}. Proceedings of the
  IEEE International Conference on Computer Vision  (2017)

\end{thebibliography}
\end{document}